\documentclass{article}
\usepackage{amssymb,amsmath,dsfont,alife10}
\usepackage[dvips]{psfrag}
\DeclareMathOperator{\Tr}{Tr}
\DeclareMathOperator{\e}{e}

\newcommand{\dd}{\text{d}}
\newcommand{\beq}{\begin{eqnarray}}
\newcommand{\eeq}{\end{eqnarray}}
\newcommand{\beqn}{\begin{eqnarray*}}
\newcommand{\eeqn}{\end{eqnarray*}}
\newcommand{\expt}[1]{\langle{#1}\rangle}

\begin{document}

\title{Quasi-Species and Aggregate Dynamics}
\author{Anders Eriksson$^1$, Olof G\"{o}rnerup$^1$, Martin Nilsson Jacobi$^{1,\dagger}$ and Steen Rasmussen$^{2,3}$\\
$^1$ Complex Systems Group, Chalmers University of Technology, G\"{o}teborg, SE-412 96, Sweden\\
$^2$ Self-Organizing Systems, EES-6, Los Alamos National Laboratory, NM 87545, USA\\
$^2$ University of Copenhagen, IMBG, DK-2200, Denmark\\
$^\dagger$ Corresponding author: mjacobi@chalmers.se
}

\maketitle
 
 % ---------------------------------------------------------------------------------------------
 % ---------------------------------------------------------------------------------------------
 % ---------------------------------------------------------------------------------------------
 % ---------------------------------------------------------------------------------------------
 % ---------------------------------------------------------------------------------------------
 % ---------------------------------------------------------------------------------------------

\begin{abstract}
At an early stage in pre-biotic evolution, groups of replicating molecules must coordinate their reproduction to form aggregated units of selection. Mechanisms that enable this to occur are currently not well understood. In this paper we introduce a deterministic model of primitive replicating aggregates, proto-organisms, that host populations of replicating information carrying molecules. Some of the molecules promote the reproduction of the proto-organism at the cost of their individual replication rate. A situation resembling that of group selection arises.  We derive and analytically solve a  partial differential equation that describes the system. We find that the relative prevalence of fast and slow replicators is determined by the relative strength of selection at the aggregate level to the selection strength at the molecular level. The analysis is concluded by a preliminary treatment of finite population size effects.
\end{abstract}

 % ---------------------------------------------------------------------------------------------
 % ---------------------------------------------------------------------------------------------
 % ---------------------------------------------------------------------------------------------

\section{Introduction}

In primitive organisms without central control of genome replication, a conflict between selfishly reproducing genes and genes useful for the replication of the whole organism may occur. This raises the question of how, and when, the organism as a whole can be viewed as a unit of selection. This is a necessary condition for such systems to  evolve into contemporary organisms, with a well-defined separation between the genotype and phenotype, and a coordinated replication. 

We study the evolutionary dynamics of systems consisting of self-assembling container aggregates that contain populations of self-replicating information carrying molecules -- proto-genes. The aggregates can be viewed as primitive proto-organisms, each with a genome consisting of an evolving population of proto-genes. The aggregates grow by successively incorporating new building blocks. Eventually tyhey become unstable and spontaneously divide, whereby a replication of the proto-organism has occurred. The production of new building blocks, e.g. amphiphilic polymers, is catalyzed by the proto-genes, e.g. through an electron charge transfer process. A strain's ability to self-replicate and its chemical properties critical to the growth of the aggregate are assumed to be uncorrelated. Certain strains of proto-genes are efficient as self-replicators, whereas other strains are more active in the production of new building blocks, and thereby contribute more to the reproduction of the container. The evolution of the system as a whole is then characterized by a conflict reminiscent of group selection. 

What conditions enable co-existence of selfish genomes and locally suppressed genomes whose presence are advantageous to the population they are members of,  or, in broader terms; what conditions allow a trade-off between local reproduction of individual sequences and global reproduction of the proto-containers that enclose them?

\section{Background}

\subsection{The quasi-species model}

As the quasi-species framework serves as a basis for the current container growth model, the former will now briefly be introduced. The quasi-species model was originally formulated by Eigen \cite{Eigen71} as a way to describe and analyse pre-biotic molecular replicator dynamics. Constituted by bit-strings of finite length, the individuals represent sequences of elementary building blocks or bases that are considered to have given characteristic traits that determine their expected number of offspring per time unit.  As a simple yet illuminating case, a single peak fitness landscape is often assumed. That is, all individuals are assigned an equal ability to reproduce, except for one---the master sequence---which is given a higher fitness. In contrast to the selective pressure, variation is implied by a limited accuracy in the asexual copying process from parent to offspring (i.e. mutations). 

Let  $x_k$ denote the relative frequency of individual $k$. The replicator dynamics of the population is then described by the rate equations
\begin{equation}
\dot{x}_k=\sum_l {Q_k^l a_l x_l} - \bar{e}x_k,
\label{qs_rate_eq}
\end{equation}
where $a_l$ is the fitness of individual $l$, $Q_k^l$ is the probability that reproduction of individual $i$ gives individual $k$ as offspring, and where $\bar{e} = a^T x = \sum_l{a_l x_l} $ is the average fitness of the population. The second term, $\bar{e}x_k$, ensures normalised concentrations.

Given large sequence lengths $\nu$ and a low mutation rate $\mu$, a useful approximation is possible by acknowledging that there is a low probability of mutating from a background sequence (that is, any sequence not being a master sequence) onto a master sequence \cite{NS89}. When employing this approximation, the population is considered to consist of two types, master- and background sequences, and the time dynamics is reduced to 
\begin{eqnarray}\label{eq:quasi_species_1}
\dot{\xi} & = & ( A -1 ) \xi  \left( \frac{Q A -1}{A-1} - \xi \right) ,
\label{qs_rate_no_bm_eq}
\end{eqnarray}
where $\xi$ is the concentration of master sequences,  $A=a_0$ is the fitness advantage of the master sequence, and the copying fidelity $Q$ is the probability that there are no mutations during a replication event. The background sequences all have fitness $1$, so the factor $A-1$ is the relative fitness advantage of the master sequence. 

The most important result from this model concerns the existence of a sharp lower limit to the copying fidelity---the error threshold---below which no information can be preserved in the population by means of the selective pressure \cite{Eigen71}.

\subsection{Eigen's paradox}

An implication of the error threshold is that large early molecular replicators---with lengths of, say, RNA viruses of today---had to reproduce with very high accuracy. Due to this requirement, specialised enzymes had to be utilised in order to correct for imposed mutations. However, these enzymes, in turn, could only be encoded by long nucleotide sequences. That is, large sequences required enzymes that required large sequences. In order to resolve this recursive problem, Eigen proposed the hyper-cycle; a mechanism with which a set of sequences cooperatively by means of auto-catalysis share the burden of information carriage \cite{Eigen77}. These constructions, though, are presumed to be highly vulnerable to parasites---i.e. molecules that benefit from catalytic support, although not contributing to the auto-catalytic circle---and does therefore not serve as a solution to the information storage dilemma in a harsh pre-biotic environment.

\subsection{Group selection}

An alternative architectural principle that in a more stable manner would allow for cooperation among information carriers---organised in hyper-cycles or not---is to form compartments. When realised, a compartment and the template molecules that it encloses may under evolution be viewed as a unit of selection whose absolute fitness is determined by the composition of its contents. 

The situation described constitutes group selection as originally studied by Wright in his Island model of spatially isolated local and macroscopic populations \cite{Wright31}. In a similar setting, \cite{Levins70}, and later \cite{Boorman80} study extinction and re-colonisation of- and by locally evolving populations.
At a smaller scale, \cite{Szathmary87,MaynardMajorTrans} analyse group selection on the level of replicative molecules. In their stochastic corrector model, small compartments encapsulate replicators of two given types, fast and slow, where the latter benefit the survival of the group and the former does not. Given that the groups consist of few molecules, thus implying a high degree of stochasticity in the system, it is shown numerically that, under certain conditions, there exists  a stable global polymorphism of fast and slow replicators.
Similarly, \cite{Alves2000} adapts Wright's island model to the domain of molecular replicators. Again, two template types, fast and slow, enclosed in finite---although not necessarily small---compartments are considered. In consistency with previously mentioned work, parameter regions that enable stable coexistence of the two types are found. In more recent work \cite{Fontanari05}, the above model is generalised to concern up to four (as limited by numerical constraints) different template types, where vesicles containing populations with high degrees of multitude are favoured. The dynamics is evaluated by numerically iterating a set of recursion equations, whereof the regions of the model's parameter space that enable coexistence of up to the four template types are identified.

\section{The full population dynamics}
 
\begin{figure}[t]
\begin{center}
\includegraphics[width=3in]{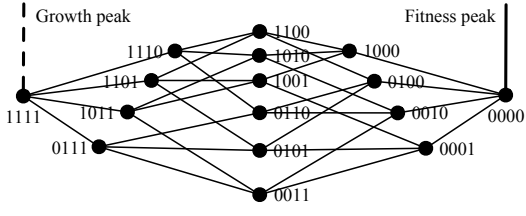}
\caption{Sequence space for $\nu=4$ with a fitness peak at the master sequence $0000$ and a aggregate growth peak at the slow replicator $1111$.}
\label{fitness_landscape}
\end{center}
\end{figure}

Consider a population of proto-containers, where each container hosts a population of individuals as formulated in the quasi-species framework. Since the populations of the separate containers are isolated from each other, each population evolves individually. In accordance with the original quasi-species setting, there is one master sequence with a higher reproduction rate. However, there is also another sequence---being the one that is furthest away from the master sequence in terms of Hamming distance---that promotes the growth of the whole container, Fig. \ref{fitness_landscape}. At a certain size, the container spontaneously divides. This constitutes a replication of the proto-organism. Since the proto-containers are subject to selection, they in turn---like the individual sequence populations---compete for maintenance.

The slow replicators that promotes the growth of the container is not favoured in the populations due to local domination of the master sequence. On the scale of the containers, though, the slow replicator is presumed to have an advantage since it enhances the fitness of its host container. 

We assume that the growth of an aggregate is directly determined by its internal concentration through some function $\phi(x)$. Let $\psi(t,x)$ denote the relative concentration of aggregates that  contain a population of information molecules with concentration vector $x$ at time $t$. We use a standard argument, see~\cite{Boorman80} for details, for the flux in the non-normalised concentration density $\widetilde{ \psi} (t,x)$ in a volume $V$ in $x$-space
\begin{eqnarray*}
	\partial_t  \int_V \dd x\, \widetilde{\psi} (t,x) & = & \mbox{in/out-flux} + \mbox{production} \\
	& = & -\ \int_S \dd S \cdot \dot{x}\, \widetilde{\psi}(t,x) + \int_V \dd x \, \phi(x)\,\widetilde{\psi} (t,x) \\
	& = &  \int_V \dd x \left( -\nabla \cdot \left[ \dot{x} \, \widetilde{\psi}(t,x) \right] + \phi (x)\,\widetilde{\psi}(t,x) \right)
\end{eqnarray*}
where $S$ is the surface enclosing the volume $V$, and $\dd S$ is a vector valued surface element pointing in the direction normal to the surface. Since the volume is arbitrary, the continuity (Euler-) equation for $\psi(t,x)$ reads
\begin{eqnarray}
	\partial_t \psi(t , x)  & = &   - \nabla \cdot \left[ \dot{x} \, \psi (t,x) \right] + \phi (x) \, \psi (t,x)  -  \nonumber \\ 			
	&& \expt{\phi , \psi} \, \psi (t,x) , 
	\label{aggr_pop_dyn_1}
\end{eqnarray}
where, just as in the regular quasi-species equation, the scalar product $\expt{\phi , \psi} ( t ) = \int \dd x\,  \phi (x) \, \psi(t,x)$ is used to normalise the relative concentrations so that $\int \dd x\, \psi (t,x) =1$ for all $t$. We use the explicit form of $\dot{x}$ given in (\ref{qs_rate_eq}) to rewrite (\ref{aggr_pop_dyn_1}) as
\begin{eqnarray}
	\partial _t \psi (t , x) & = & - x^T M ^T \nabla  \psi ( t,x) + a ^T X  \nabla \psi  (t,x)  \nonumber \\ 
	& & + \widetilde{\phi} (x) \psi (t,x)  -\expt{\phi , \psi}   \psi (t,x) ,
	\label{aggr_pop_dyn_2}
\end{eqnarray}
the matrix $X$ is defied as  $X _{i j} = x _i x _j$, and the effective growth is defined as 
\begin{eqnarray*}
	\widetilde{\phi} (x) & = & \phi (x) + (N+1)  a ^T x  - \Tr (M) ,
\end{eqnarray*}
where $N$ is the number of different information molecules, i.e., $x _i$ $i = 1 , \dots , N$, and $\Tr (M) = \sum _i M _{i i }$.  We note that (\ref{aggr_pop_dyn_2}) is only nonlinear in the re-normalisation term. A transformation 
\begin{eqnarray*}
	\widetilde{\psi} (t,x) & = & \exp\!\left[ \int^t \dd s\, \expt{\phi , \psi} ( s ) \right] \psi (t,x)
\end{eqnarray*}	
gives a linear equation in the new, non-normalised, ``distribution"
\begin{eqnarray}
	\partial _t \widetilde{\psi} (t , x)  =  - x^T M ^T \nabla  \widetilde{\psi} ( t,x) + a ^T X  \nabla \widetilde{\psi}  (t,x)    .
	\label{aggr_pop_dyn_3}
\end{eqnarray}
The normalised distribution is given by
\begin{eqnarray*}
	\psi (t ,x ) & = & \frac{\widetilde{\psi} ( t , x )}{ \int \dd \xi\, \widetilde{ \psi} (t, \xi ) }
\end{eqnarray*}

% ---------------------------------------------------------------------------------------------
% ---------------------------------------------------------------------------------------------
% ---------------------------------------------------------------------------------------------

\section{A two-state approximation}

The full version of the coupled population dynamics presented in the previous section is hard to analyse analytically. In this section we present a simplified set of equations that readily allows analytic treatment. The main idea is to use the same ``no back mutations" approximation of the quasi-species dynamics as was used to derive (\ref{qs_rate_no_bm_eq}). 

%---------------------------------------------------------------------------------------------------
\begin{figure}
	% \vspace{-2mm}
	\begin{center}
	\psfrag{x}[][]{$\xi(t)$}  \psfrag{t}[b][]{$t$}
	\includegraphics[width=7cm]{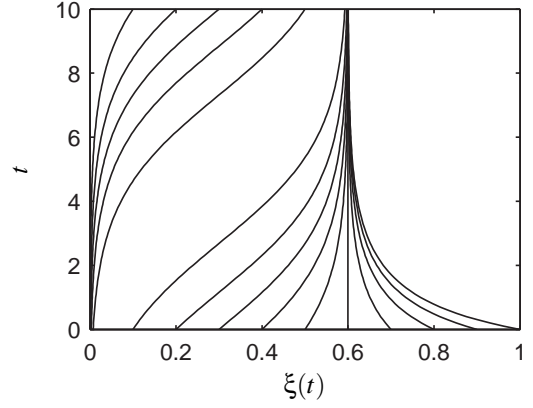}\\[-2mm]
\caption{\label{fig:qs_plot}
Trajectories $\xi(t)$ in the quasi-species dynamics ($\mu = 0.2$ and $A = 2$). All trajectories converge to $\xi = \xi_*$ (here, $\xi_* = 0.6$); as consequence, all populations with $\xi(t) < \xi_*$ for $t \gg 1$ must have $\xi(0) \ll 1$.
} 
	\end{center}
\end{figure}
%---------------------------------------------------------------------------------------------------

We make a two state approximation, assuming that the main dynamics in the model is captured only by the relative concentration of fast replicators $\xi $; all background sequences $1 - \xi $ are assumed to be beneficial for the aggregate growth. Using (\ref{qs_rate_no_bm_eq}), we can write (\ref{aggr_pop_dyn_3}) as
\begin{eqnarray}
	\partial _t \widetilde{\psi} (t, \xi  )  & = &  - \partial _{\xi } \,  \xi \, \left(  \xi_*  - \xi  \right)  \,  \widetilde{\psi} (t , \xi )  \nonumber \\
	& & + \gamma \, (1-\xi  ) \,  \widetilde{\psi} ( t , \xi  ) ,
\label{two_state_eq}
\end{eqnarray}
where $\xi _* = (A Q -1)/(A-1)$ is the equilibrium of the master sequence population dynamics and the parameter $\gamma$ is defined as 
\begin{eqnarray*}
	\gamma & = & R/(A-1) .
\end{eqnarray*}
Eq.~\ref{two_state_eq} can be solved analytically:
\begin{eqnarray}
	\widetilde{\psi} (t , \xi  )  =  \frac{\xi  ^{\alpha}}{| \xi _* -  \xi |^{\beta}} F\!\left(  \frac{\xi _* - \xi  }{\xi } \, \e^{\xi _* t}   \right)  ,
\label{solution1}
\end{eqnarray}
where $F$ is a function determined by the initial distribution, and the parameters are defined as $\alpha  =  \gamma  \xi _* ^{-1} - 1$ and $\beta  = 1 + \gamma ( \xi _* ^{-1} -1 )$. We now solve for $F$ as a function of the initial distribution $\psi_0(\xi) = \widetilde{\psi}( 0 ,  \xi  ) = \psi( 0 , \xi  )$
 \begin{eqnarray*}
	 F\!\left( \frac{ \xi _* -  \xi  }{\xi } \right)  & = & \frac{ |\xi _* - \xi |^\beta}{\xi^{\alpha}} \psi  _0 ( \xi  ) .
\end{eqnarray*}
Defining $\eta = (\xi _* -  \xi )/\xi$ gives
\begin{eqnarray}
	F ( \eta )   =    \xi _*  ^{\beta - \alpha} \, (1 + \eta) ^{ \alpha - \beta} \,  | \eta |^\beta \,  \psi _0 \!\left( \frac{\xi _*}{1 + \eta} \right) .
\label{dummy}
\end{eqnarray}
Substituting (\ref{dummy}) back into (\ref{solution1}) and, where it is convenient, using the relation $\alpha - \beta = \gamma -2$, gives the final solution
%
%\begin{eqnarray}
%	\widetilde{\psi} (t , \xi  )  & = & \exp ( \beta \xi _* t ) \left( \frac{}{} \xi + \exp ( t \xi _* ) ( \xi _* - \xi ) \right) ^{\gamma -2  } \nonumber \\
%	& & \times\ \psi _0\!\left( \frac{ \xi _* }{1 + \xi ^{-1} ( \xi _* - \xi ) \exp (\xi _* t) } \right) 
%\label{solution2}
%\end{eqnarray}
\begin{multline}\label{solution2}
	\widetilde{\psi} (t , \xi  )  =   \frac{ \xi _* ^{2- \gamma} \e^{\beta \xi_* t}} 
	{\left[ \xi + ( \xi _* - \xi )\,\e^{\xi_* t} \right] ^{2 - \gamma  }}  \; % \times \\ 	 	
	\psi _0\!\!\left( \frac{ \xi _* \xi }{\xi +  ( \xi _* - \xi )\, \e^{\xi_* t} } \right) 
	% \text{H}\!\!\left( \frac{\xi_* \, \e^{\xi_* t}}{\e^{\xi_* t} - 1 + \xi_*} - \xi \right)
\end{multline}
when $0 \le \xi \le \xi_* /[ 1 - (1 - \xi_*)\,\e^{-\xi_* t} ]$ and, as a consequence  of $\psi _0 ( \zeta ) = 0$ when $\zeta >1$, zero otherwise.
%\beq
% x + e s - e x > 0 <=> (e - 1) x > e s  <=> x > s e/(e - 1) 
% x s /(x + (s - x)e) < 1
% s x < x + e s - e x
% (e - 1 + s) x < e s
% x < s e/(e -1 + s) < s e/(e -1) 
%	\xi &\le& \xi_* \left[ 1 - (1 - \xi_*)\,\e^{-\xi_* t} \right]^{-1} 
%\eeq
%and zero else.

For large $t$, (\ref{solution2}) is significantly simplified. We need to consider two regions separately. Close to the singularity $\xi  = \xi _*$ the solution behaves different than elsewhere. 
\begin{eqnarray*}
	\widetilde{\psi} (t , \xi  )  & = &  \exp (  \beta  \xi _*  t )   \psi _0  \left( \xi _* \right) 
\end{eqnarray*}
when $| \xi   - \xi _* | < \exp ( - \xi _* t)$, whereas
\begin{eqnarray*}
	\widetilde{\psi} (t , \xi  )  & =& \exp ( \alpha \xi _*  t ) \left( \frac{ \xi _* - \xi}{\xi _*} \right) ^{\gamma - 2 } \psi _0 (0)  
\end{eqnarray*} 
otherwise. The weight of the population located around $\xi  = \xi _*$, i.e., the fast replicators,  grows like $W _e \sim \exp ( \beta \xi _* t )  \exp ( - \xi _* t)$, while the rest of the population has weight $W_a \sim \exp ( \alpha \xi _* t)$. The conclusion is that if $\gamma <   1$, then the entire population will be concentrated in an infinitesimal surrounding of $\xi  = \xi _*$, i.e., the fast replicators dominate the population. 

If $\gamma>1$, the distribution is given by
\begin{eqnarray}
	\psi _{stat} (  \xi  ) & = & Z ^{-1}\, ( \xi _* -  \xi )^{\gamma -2} ,
\label{solution_dist}
\end{eqnarray}
where $Z$ is just a normalisation factor. We note that the weight, i.e., the integral of the distribution, close to $\xi  = \xi _*$ converges when $\gamma >1$. Eq.~\ref{solution_dist} also changes behaviour when $\gamma -2 = 0$. Then the population changes from being dominated by fast replicators to a situation when the slow replicators dominate. The weight of the distribution shifts towards $\xi =0$. 

Our analysis of the asymptotic behaviour assumes that the initial distribution is smooth. Most importantly we assume that $\psi _0 ( \xi ) $ is regular at $\xi = 0$. It is clear already from (\ref{two_state_eq}) that $\xi =0$ is a fixed point of the dynamics. If, for example, the initial distribution has a delta function at $\xi = 0$, i.e. $\psi _0 (\xi ) = b \delta ( \xi ) + \chi ( \xi )$ where $\chi$ is a smooth function, then the weight of the distribution around $\xi =0$ grows as $\exp ( \gamma t)$ relative to the weight of the rest of the interval. Clearly then, the asymptotic distribution is $\psi _{stat} ( \xi ) = 2 \delta ( \xi )$, i.e. there are only slow replicators in the population. The conclusion is that the stationary distribution is completely dominated by slow replicators if there initially exists containers with no fast replicators. If the population within an aggregate is finite (see the discussion in the next section), this situation is not be unlikely. If fact, this observation is used in the stochastic corrector model of group selection \cite{Szathmary87}. We conclude the analysis by reviewing the three different cases, remembering the definition $\gamma = R/(A -1)$:
\begin{description}
\item[I.] $R < 2( A-1)$, $\psi _0 (\xi )$ regular at $\xi = 0$: The fast replicators dominate the total population.
\item[II.] $R > 2( A-1)$, $\psi _0 (\xi )$ regular at $\xi = 0$: The slow replicators dominate the total population but some fast replicators still exist.
\item[III.] $\int _0 ^{\epsilon} \dd \xi \psi _0 (\xi ) \geq \delta > 0$  when $\epsilon \rightarrow 0^+$: The slow replicators completely dominate the total population.
\end{description}
Note that the condition $R \gtrless 2 (A-1)$ is independent of the copying fidelity $Q$. This can be understood from (\ref{qs_rate_no_bm_eq}) where we see that $A-1$ measures the rate of convergence towards the stationary distribution $\xi _*$.
%---------------------------------------------------------------------------------------------------
%---------------------------------------------------------------------------------------------------
%---------------------------------------------------------------------------------------------------

\section{Finite size effects}

%The results in the previous section were derived in the infinite population size limit. In this section, we introduce and analyse a variant of the model with a finite number $n$ of containers. The internal dynamics of the number of information molecules is determinstic as before, but container division occurs according to a stochastic process. In the limit of $n \rightarrow \infty$, $\overline{\psi}(t,x)$, the distribution $\psi(t,x)$ averaged over all realisations of the stochastic dynamics, approaches the solution to (\ref{two_state_eq}). 

The results in the previous section were derived in the infinite population size limit. In this section, we introduce and analyse a variant of the model with a finite number $n$ of containers. 
%The internal dynamics of the number of information molecules is determinstic as before, but container division occurs according to a stochastic process. 
Each container $i$ in the population is characterised by the fraction $\xi_i$ of information molecules in the container that replicate efficiently. As before, it is assumed that within each container, the fraction $x_i$ evolves according to the quasi-species dynamics in (\ref{eq:quasi_species_1}). Container $i$ divides according to an in-homogenous Poisson process with (instantaneous) rate $\gamma\,[1 - \xi_i(t)]$, corresponding to the second term in (\ref{two_state_eq}). We assume that the container divides in two equal halves, with identical composition of information molecules.

In our  simulations, we take the initial population, $\xi_i(0)$, to be independent uniformly distributed random numbers. The time steps in the simulations consist of two parts. First, the value $\xi_i$ of each container $i$ is updated as
\beq
	\xi_i(t+\delta t) &=& \xi_i(t) + \delta  t\, \xi_i(t) \, [\, \xi_* - \xi_i(t) \,] .
\eeq
Second, each container is tested for division. With probability $1 - \exp(-\gamma \delta t(1-\xi_i))$ we copy container $i$ to a randomly chosen container in the population. This allows for the correct rate of division for each container, while the number for containers in the population is kept constant. When $\gamma \delta t \ll 1$, the probability of division is approximately $\gamma \delta t(1-\xi_i)$. Throughout this section, we use $\delta t = 0.01$ and $n = 10^5$.

The model behaves smoothly in $\mu$ as $\mu \rightarrow 0$, so there is very little difference between the dynamics for, e.g., $\mu = 10^{-3}$ and $\mu = 0$. Hence, in order to simplify the analysis we take $\mu = 0$ in all simulations. The theoretical predictions are then
\beq\label{eq:psi_theory}
	\psi(t,\xi) &=& \frac{(\gamma - 1)(1 - \e^{-t})}{1 - \e^{-(\gamma - 1)t}} \left(1-\xi + \xi\,\e^{-t} \right)^{\gamma-2}
\eeq
for the distribution of $\xi$ and
\beq\label{eq:expt_xi_theory}
	\expt{\xi}(t) &=& \frac{1}{\gamma} \left[  \frac{\gamma + (1 - \gamma)\,\e^{-t}}{1 - \e^{-t}} +
					     \frac{1 - \gamma}{1 - \e^{-(\gamma - 1)t}} \right]
\eeq
for the expected value of $\xi$.

\begin{figure}
	% \vspace{-2mm}
	\begin{center}
	\psfrag{xlabel}[t][]{$t$}  \psfrag{ylabel}[b][]{Expected value of $\xi$}
	\includegraphics[width=7cm]{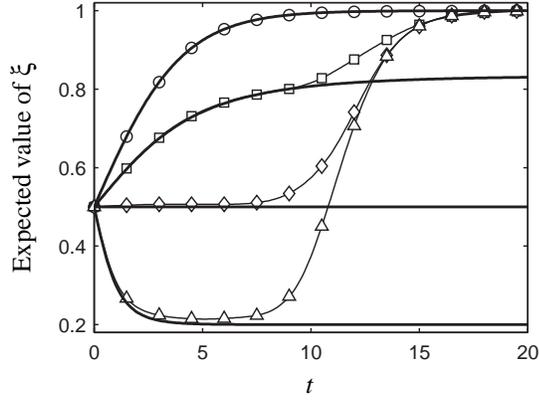}\\[-2mm]
\caption{\label{fig:avg}
Development of the expected value of $\xi$. Simulation results, averaged over 100 runs, are shown as lines decorated with symbols for $\gamma = 0.5$ (circles), $\gamma = 1.2$ (boxes), $\gamma = 2$ (diamonds), and $\gamma = 5$ (triangles). Also shown is (\ref{eq:expt_xi_theory}) for each value of $\gamma$ (thick lines).
} 
	\end{center}
\end{figure}

%Figures \ref{fig:avg}--\ref{fig:distr} illustrates the simulation results. 
In Fig.~\ref{fig:avg}, we show the expected value of $\xi$ as a function of time for parameter values corresponding to the three regions, $0 < \gamma < 1$, $1 < \gamma < 2$ and $\gamma > 2$, and the boarder case $\gamma = 2$. Also shown is the theoretical prediction (\ref{eq:expt_xi_theory}). For $t \lesssim 8$, the simulation results and theory agree, but for larger $t$ all simulation curves rise to $\expt{\xi} = 1$.

%When $0 < \gamma < 1$,  containers with large fraction of fast-replicating information molecules dominate the population for all times, as expected from the analysis in the previous section. See Fig.~\ref{fig:avg} for the expected value of $\xi$ as a function of time, for $\gamma = 0.5$ (dotted line). Also shown is the theoretical prediction of (\ref{eq:expt_xi_theory}).

%---------------------------------------------------------------------------------------------------
%---------------------------------------------------------------------------------------------------

%In the region $1 < \gamma < 2$, we still expect the fast-replicating molecules to take over eventually, but the results from the simulations for $\gamma = 1.2$ (see Fig.~\ref{fig:avg}) show a increase around $t \approx 5$ compared to the curve predicted by the theory. Also show is the curve for the limiting case $\gamma = 2$ (see Fig.~\ref{fig:avg}). For this case, theory predicts that $\expt{\xi}(t)$ is constant in time. Again, deviation from theory occurs at $t \approx 5$.

%When $\gamma > 2$, the theory predicts that the population will converge a distribution $\propto (1 - \xi)^{\gamma - 2}$ as $t$ grows. As can be seen for $\gamma = 5$ in Fig.~\ref{fig:avg}, this prediction is consistent with the time evolution of $\expt{\xi}(t)$ for $t \lesssim 6$.

We identify two sources of the differences between the simulated model and the infinite population size model in the previous section. First, in the simulations, the production of new containers by division is a stochastic process -- as opposed to the deterministic growth term in (\ref{two_state_eq}). Hence, the deterministic time evolution of $\psi(t,\xi)$ in (\ref{two_state_eq}) acquires a noise term, which in turn causes  $\psi(t,\xi)$ to be smeared out.

Second, in addition to the effects of stochastic division of containers, we have the consequences of small differences in the initial distribution from one run to the next. When $t$ is small, it is still a good approximation to say that $\psi_0(\xi)$ is a smooth function. However, when $t$ is large the distribution $\psi(t,\xi)$ depends on the initial distribution only in a small interval close to $\xi = 0$ (see Fig.~\ref{fig:qs_plot}). Hence, in a finite population we can no longer assume that $\psi_0(\xi)$ is smooth when considered at such a small scale: at long times the dynamics will then be dominated by the containers with the smallest $\xi(0)$. Note how this directly relates to the third case discussed at the end of the last section. 
%Let $\xi_\text{min}$ be the minimum inital $\xi$ in the population. Since we take $\xi_i(0)$ to be indipendent uniformly distributed random numbers, the distribution of $\xi_\text{min}$ is 
%\beqn\label{eq:distr_xi_min}
%	f(\epsilon) &=& n\,(1 - \epsilon)^{n-1} \approx (n-1)\,\e^{-(n-1)\epsilon}.
%\eeqn
%The expected value of $\xi_\text{min}$ is $1/(n+1)$.

The question is now: which of these two effects is the dominant cause for the deviations observed from the theoretical predictions? In order to answer this question, suppose the initial distribution is uniform on the interval $[\epsilon,1]$. According to (\ref{solution2}) the distribution is then
\begin{multline}\label{eq:psi_theory2}
	\psi_\epsilon(t,\xi) \  =\  \frac{ (\gamma - 1)(1 - \e^{-t}) }{ \left(1 - \epsilon + \epsilon\,\e^{t} \right)^{1 - \gamma} - \e^{-(\gamma - 1)t} }\, 
	 (1 - \xi + \xi\,\e^{-t} )^{\gamma-2} \hspace{-2mm}
\end{multline}
on the interval $[\xi_\text{min}(t), 1]$ and zero outside, where
\beqn
	\xi_\text{min}(t) &=& \epsilon/ [\, \epsilon + (1 - \epsilon)\,\e^{-t} \, ]  \, .
\eeqn
The corresponding expected value is
\begin{multline}\label{eq:expt_xi2}
	\expt{\xi_\epsilon}(t) \ =\  \frac{1}{\gamma} \, \frac{\gamma + (1 - \gamma)\,\e^{-t}}{1 - \e^{-t}}\ + \\
				  +\  \frac{1}{\gamma} \,  \frac{(1-\gamma) (1-\epsilon)\,\e^{-t}}{\epsilon + (1 - \epsilon)\,\e^{-t} - [\, \epsilon + (1 - \epsilon)\,\e^{-t} \,]^{\gamma} } .
\end{multline}
%---------------------------------------------------------------------------------------------------
\begin{figure}[t]
	% \vspace{-2mm}
	\begin{center}
	\psfrag{xlabel}[t][]{$t$}  \psfrag{ylabel}[b][]{Expected value of $\xi$}
	\includegraphics[width=7cm]{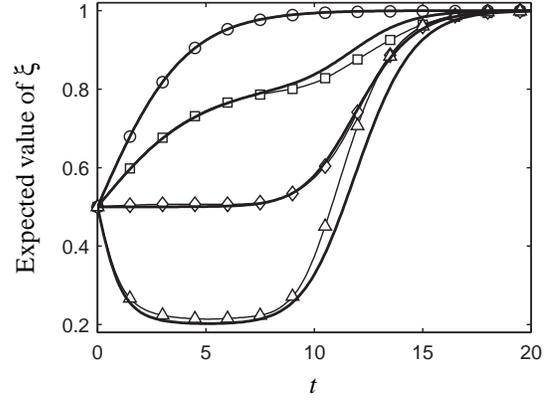}\\[-2mm]
\caption{\label{fig:avg2}
Expected value of $\xi$ from simulations and from (\ref{eq:expt_xi2}), averaged over $\xi_\text{min}$.
Simulation results, averaged over 100 runs, are shown as lines decorated with symbols for $\gamma = 0.5$ (circles), $\gamma = 1.2$ (boxes), $\gamma = 2$ (diamonds), and $\gamma = 5$ (triangles). 
} 
	\end{center}
\end{figure}
%---------------------------------------------------------------------------------------------------

\begin{figure}[t]
	% \vspace{-2mm}
	\begin{center}
	\begin{tabular}{l@{}l}
		\psfrag{xlabel}{}  \psfrag{ylabel}[b][]{$\psi$} \psfrag{cval}[rt][]{\shortstack{$\gamma = 0.5$\\$t = 0.5$}}
		\includegraphics[width=4cm]{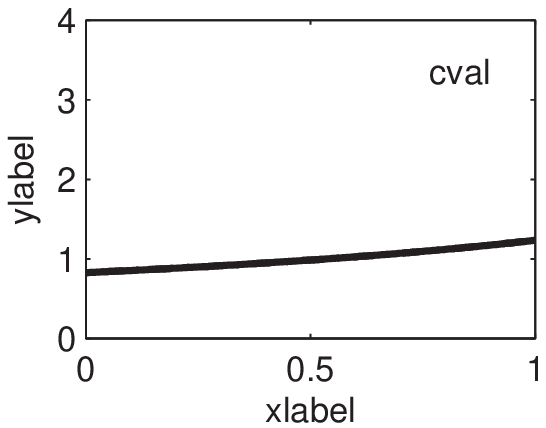} &
		\psfrag{xlabel}{} \psfrag{ylabel}{}  \psfrag{cval}[rt][]{\shortstack{$\gamma = 5$\\$t = 0.5$}}
		\includegraphics[width=4cm]{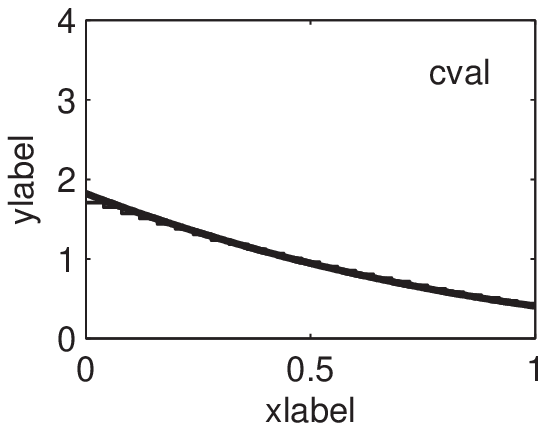} \\[-3mm]
		%------------------------------------------------------------------------------------------------
		%
		\psfrag{xlabel}{} \psfrag{ylabel}[b][]{$\psi$} \psfrag{cval}[rt][]{\shortstack{$\gamma = 0.5$\\$t = 4$}}
		\includegraphics[width=4cm]{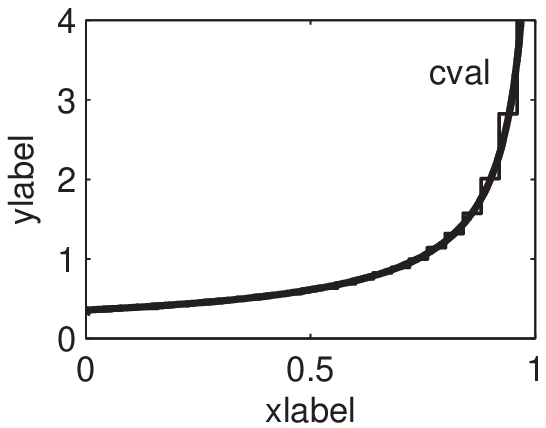} &
		\psfrag{xlabel}{} \psfrag{ylabel}{} \psfrag{cval}[rt][]{\shortstack{$\gamma = 5$\\$t = 4$}}
		\includegraphics[width=4cm]{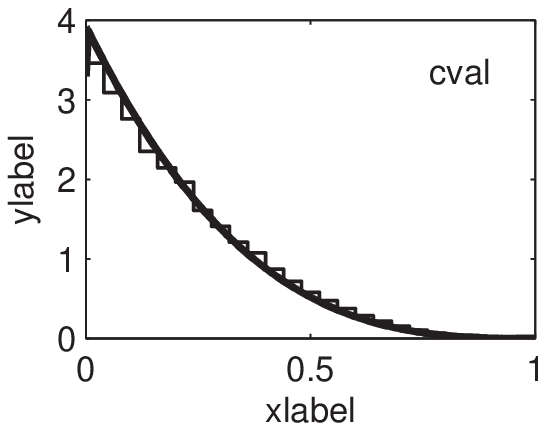} \\[-3mm]
		%------------------------------------------------------------------------------------------------
		%
		\psfrag{xlabel}{} \psfrag{ylabel}[b][]{$\psi$} \psfrag{cval}[rt][]{\shortstack{$\gamma = 0.5$\\$t = 10$}}
		\includegraphics[width=4cm]{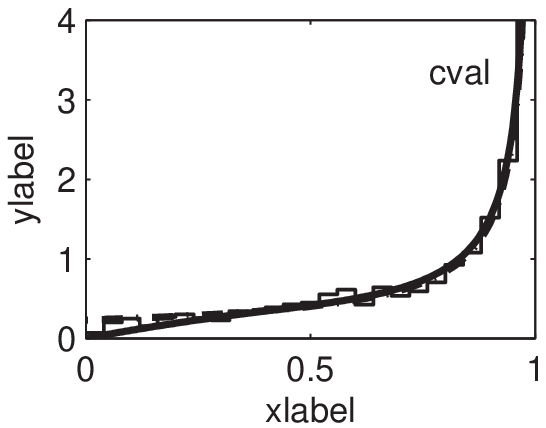} & 
		\psfrag{xlabel}{}  \psfrag{ylabel}{} \psfrag{cval}[rt][]{\shortstack{$\gamma = 5$\\$t = 10$}}
		\includegraphics[width=4cm]{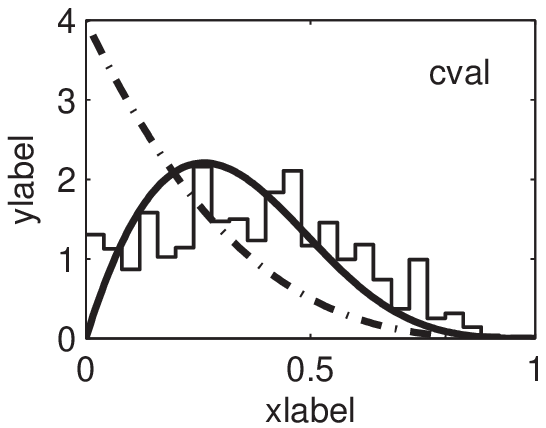} \\[-3mm] 
		%------------------------------------------------------------------------------------------------
		%
		\psfrag{xlabel}[t][]{$\xi$} \psfrag{ylabel}[b][]{$\psi$} \psfrag{cval}[rt][]{\shortstack{$\gamma = 0.5$\\$t = 14$}}
		\includegraphics[width=4cm]{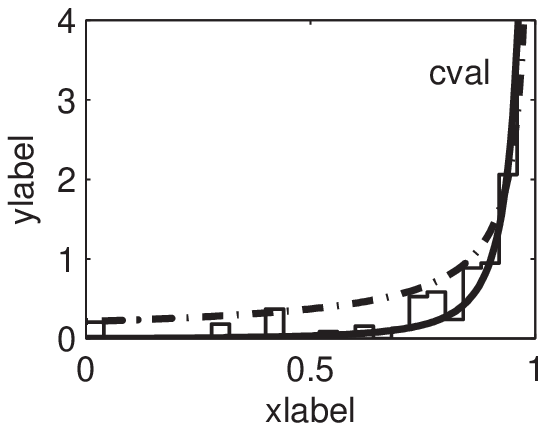} &
		\psfrag{xlabel}[t][]{$\xi$} \psfrag{ylabel}{} \psfrag{cval}[rt][]{\shortstack{$\gamma = 5$\\$t = 14$}}
		\includegraphics[width=4cm]{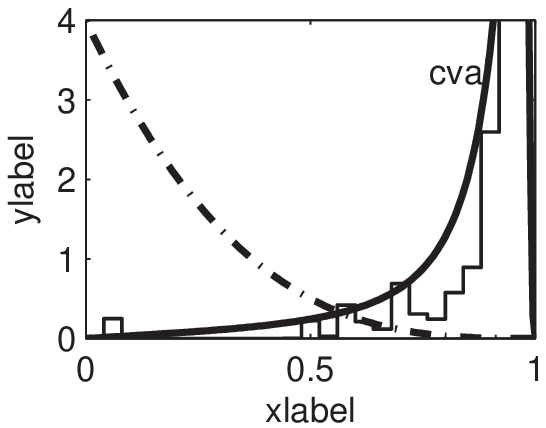}	
	\end{tabular}
\caption{\label{fig:distr}
Distribution of $\xi$ for $\gamma = 1.2$ and $\gamma=5$ at different times $t$. Average histograms from 100 simulations is shown as a stair-case plot. Also shown is $\psi(t,\xi)$ from (\ref{eq:psi_theory}) (dashed and dotted line) and $\psi_\epsilon(t,\xi)$ averaged over the distribution of $\epsilon$ (solid line).
} 
	\end{center}
\end{figure}

In Fig.~\ref{fig:avg2} we show the expected value of $\xi$, for the same parameters as in Fig.~\ref{fig:avg}, with (\ref{eq:expt_xi2}) averaged over the distribution  $n(1 - \epsilon)^{n-1}$, corresponding to the minimum of $n$ randomly chosen uniformly distributed numbers. This model captures the transition from $\expt{\xi}$ to $1$, but cannot explain the simulation results completely: we attribute the remaining differences to the effect of stochastic fluctuations of the dynamics. Comparing the distribution of $\xi$, estimated from the simulations, to the theoretical prediction (\ref{eq:psi_theory}) and the second model (\ref{eq:psi_theory2}) (Fig.~\ref{fig:distr}) supports this conclusion.

%
%In Fig.~\ref{fig:distr} we compare the distribution of $\xi$ at different times and for two values of $\gamma$, estimated from simulations, to the theoretical prediction (\ref{eq:psi_theory}) and the second model (\ref{eq:psi_theory2}). The second model is a much better fit to the simulations, but does not capture the simulation results fully. 
%
%---------------------------------------------------------------------------------------------------

Finally, we consider under which circumstances one can expect (\ref{eq:psi_theory}) to predict the distribution of $\xi$. For $\epsilon \lesssim 10^{-2}$, $\expt{\xi_\epsilon} \approx \expt{\xi}$ when $t \lesssim \log(0.1/\epsilon)$. For larger $t$, $\expt{\xi_\epsilon}$ approaches $1$. In the limit $\epsilon \rightarrow 0$, the population converges to the stationary solution at $t \approx 4$. Hence, in order for the population to have time to approach the stationary solution before the finite size effects take over, $\epsilon^{-1}$, i.e. the number of replicators in a typical container,  must be at least 500.

 % ---------------------------------------------------------------------------------------------
 % ---------------------------------------------------------------------------------------------
 % ---------------------------------------------------------------------------------------------

\section{Conclusions}

The deterministic model is a good description of the dynamics of the system during a transient. Most importantly, the condition $R \gtrless 2 (A-1)$ decides whether the distribution is dominated by  fast or slow replicators. The long-term dynamics, on the other hand, is determined mostly by the distribution of the smallest values of $\xi$ in the inital population. The stochastic division process influence the dynamics, but has very little qualitative effect (at least for large population sizes). These are only preliminary results; the effects of finite containers, stochastic growth processes and other complications to the model need to be investigated further before a conclusive answer can be obtained.

%---------------------------------------------------------------------------------------------------
%---------------------------------------------------------------------------------------------------
%---------------------------------------------------------------------------------------------------

\section{Acknowledgements}

This work was funded by PACE (Programmable Artificial Cell Evolution), a European Integrated Project in the EU FP6-IST-FET Complex Systems Initiative. The authors would also like to thank John McCaskill for comments and discussion.

%---------------------------------------------------------------------------------------------------
%---------------------------------------------------------------------------------------------------
%---------------------------------------------------------------------------------------------------

\bibliographystyle{alife10}
\bibliography{aggr_quasi}

\begin{thebibliography}{}

\bibitem[Alves et~al., 2001]{Alves2000}
Alves, D., Campos, P.~R., Silva, A.~T., and Fontanari, J.~F. (2001).
\newblock Group selection models in prebiotic evolution.
\newblock {\em Physical Review E}, 63(1):011911--1--011911--9.

\bibitem[Boorman and Levitt, 1980]{Boorman80}
Boorman, S.~A. and Levitt, P.~R. (1980).
\newblock {\em The genetics of altruism}.
\newblock Academic Press, New York.

\bibitem[Eigen, 1971]{Eigen71}
Eigen, M. (1971).
\newblock Self-organization of matter and the evolution of biological
  macromolecules.
\newblock {\em Naturwissenschaften}, 58:465--523.

\bibitem[Eigen and Schuster, 1977]{Eigen77}
Eigen, M. and Schuster, P. (1977).
\newblock The hypercycle. {A} principle of natural self-organization. {P}art
  {A}: emergence of the hypercycle.
\newblock {\em Naturwissenschaften}, 64:541--565.

\bibitem[Fontanari et~al., 2005]{Fontanari05}
Fontanari, J.~F., Santos, M., and Szathm\'{a}ry, E. (2005).
\newblock Coexistence and error propagation in pre-biotic vesicle models: A
  group selection approach.
\newblock {\em Journal of Theoretical Biology}.
\newblock Electronic publication, ahead of print.

\bibitem[Levins, 1970]{Levins70}
Levins, R. (1970).
\newblock Extinction.
\newblock In Gerstenhaber, M., editor, {\em Some Mathematical Questions in
  Biology. Lecture Notes on Mathematics in the Life Sciences}, pages 75--107.
  The American Mathematical Society, Providence.

\bibitem[Nowak and Schuster, 1989]{NS89}
Nowak, M. and Schuster, P. (1989).
\newblock Error thresholds of replication in finite populations: Mutation
  frequencies and the onset of {M}\"uller's ratchet.
\newblock {\em Journal of Theoretical Biology}, 137:375--395.

\bibitem[Smith and Szathm\'{a}ry, 1995]{MaynardMajorTrans}
Smith, J.~M. and Szathm\'{a}ry, E. (1995).
\newblock {\em The major transitions in evolution}.
\newblock W.H. Freeman.

\bibitem[Szathm\'{a}ry and Demeter, 1987]{Szathmary87}
Szathm\'{a}ry, E. and Demeter, L. (1987).
\newblock Group selection of early replicators and the origin of life.
\newblock {\em Journal of Theoretical Biology}, 128:463--486.

\bibitem[Wright, 1931]{Wright31}
Wright, S. (1931).
\newblock Evolution in mendelian populations.
\newblock {\em Genetics}, 16:16--97.

\end{thebibliography}

\end{document}